\documentclass[%
superscriptaddress,
 amsmath,amssymb,
   twocolumn,
 aps,
 prl,
]{revtex4-1}

\usepackage[pdftex]{graphicx}
\usepackage[pdftex]{hyperref}
\usepackage{dcolumn}
\usepackage{bm}
\usepackage{color,soul}

\usepackage{relsize}
\newcommand*{\defeq}{\stackrel{\mathsmaller{\mathsf{def}}}{=}}

\newlength{\mywidth}
 \usepackage{comment}
 \usepackage[normalem]{ulem} 
\definecolor{g-blue}{rgb}{0.83,0.95,1}
\definecolor{g-yellow}{rgb}{1,1,0.7}
\definecolor{g-green}{rgb}{0.9,1,0.9}
\definecolor{green}{rgb}{0,0.6,0}
\definecolor{cyan}{rgb}{0,0.7,0.7}
\definecolor{black}{rgb}{0,0,0}
\definecolor{grey}{rgb}{0.4 ,0.4 ,0.4 }

\usepackage{bm}
\usepackage{amssymb}
\usepackage{amsfonts}
\usepackage{amsmath}
\usepackage{graphicx}
\usepackage{amsmath,bm,epsfig}

\definecolor{g-blue}{rgb}{0.83,0.95,1}
\definecolor{g-yellow}{rgb}{1,1,0.7}
\definecolor{g-green}{rgb}{0.9,1,0.9}
\definecolor{green}{rgb}{0,0.6,0}
\definecolor{cyan}{rgb}{0,0.7,0.7}
\definecolor{black}{rgb}{0,0,0}
\definecolor{grey}{rgb}{0.4 ,0.4 ,0.4 }

\newcommand{\eq}[1]{(\ref{#1})}
\newcommand{\Eq}[1]{Eq.\,(\ref{#1})}
\newcommand{\Eqs}[1]{Eqs.\,(\ref{#1})}
\newcommand{\Fig}[1]{Fig.\,\ref{#1}}
\newcommand{\B}[1]{{\bm{#1}}}
\newcommand{\C}[1]{{\mathcal{#1}}}    

\renewcommand{\sb}[1]{_{\text {#1}}}  

\renewcommand{\k}{{\bf k}}

\renewcommand{\c}{c_{\rm s}}

\newcommand\ADD[1]{{#1}} 

\renewcommand{\c}{c_{\rm s}}

\begin{document}

\title{Energy  spectrum of two-dimensional acoustic  turbulence}
\author{Adam Griffin}
\affiliation{Institut de Physique de Nice, Universit\'e C\^ote d'Azur, CNRS, Nice, France}
\author{Giorgio Krstulovic}
\email{krstulovic@oca.eu}
\affiliation{ Université Côte d'Azur, Observatoire de la Côte d'Azur, CNRS,
Laboratoire Lagrange,  Boulevard de l'Observatoire CS 34229 - F 06304 NICE Cedex 4, France.}
\author{Victor L'vov}%
\affiliation{Dept. of Chemical and Biological Physics, Weizmann Institute of Science, Rehovot 76100, Israel}%
\author{Sergey Nazarenko}%
\affiliation{Institut de Physique de Nice, Universit\'e C\^ote d'Azur, CNRS, Nice, France}%
\date{\today}
\begin{abstract} 
  We report an exact unique constant-flux power-law analytical solution of the wave kinetic equation for the turbulent energy spectrum, $E(k)=C_1 \sqrt{\varepsilon\, a c\sb s}/k$,  of acoustic waves in 2D with almost linear dispersion law, $\omega_k = c\sb s k[1+(ak)^2]$, $ ak \ll 1$.
  Here $\varepsilon$ is the energy flux over scales, and  $C_1$ is the universal constant which was found analytically. Our theory describes, for example,  acoustic turbulence in 2D Bose-Einstein condensates (BECs). The corresponding 3D counterpart of turbulent acoustic spectrum was found over half a century ago, however, due to the singularity in 2D, no solution has been obtained until now.  
  We show the spectrum  $E(k)$ is realizable in direct numerical simulations of forced-dissipated Gross-Pitaevskii equation in the presence of strong condensate. 
\end{abstract} 

\maketitle  

\noindent 

Waves in nonlinear systems interact and transfer energy along scales in a cascade process \ADD{with a constant flux}, creating an out-of-equilibrium state known as wave turbulence. When nonlinearity is   small, the Weak-Wave Turbulence  (WWT)  theory provides a mathematical description of the system \cite{ZLF,Nazarenko_2011}. The most common applications of     this theory  are capillary-gravity waves \cite{Falcon_ExperimentsSurfaceGravity_2022}, Alfv\'en waves in magnetohydrodynamics \cite{Galtier_WeakTurbulenceTheory_2000}, Langmuir waves in plasmas \cite{Zakharov_CollapseLangmuirWaves_}, inertial and internal waves in rotating stratified fluids \cite{Caillol_KineticEquationsStationary_2000,Galtier_WeakInertialwaveTurbulence_2003}, Kelvin waves in vortices \cite{Lvov_WeakTurbulenceKelvin_2010}, elastic plates \cite{during2006weak}, gravitational waves \cite{galtier2017turbulence} and density waves in Bose-Einstein condensates \cite{Dyachenko:1992aa}.

 Note, that  acoustic waves in ideal compressible fluids, one of the most common examples in Nature, are non-dispersive: their frequency  $\omega_{\B k}= c\sb s k$ is linear in  wave number $k\equiv |\B k|$. Accordingly, the resonance conditions 
\begin{equation}\label{CL}
   \omega_{\B k}= \omega_{\B k_1} + \omega_{\B k_2}\,, \quad  \B k =\B k_1+\B k_2\,, 
\end{equation}
allow for interaction of the waves with parallel wave vectors only:  $\B k \|\B k_1\|\B k_2$ \ADD{and in the same direction} \cite{Nazarenko_2011}. Therefore in the reference frame moving with the speed of sound  $c\sb s$ in the direction of   $\B k$, $\B k_1$ and $\B k_2$ all wave packets are at the rest and their overlapping  time $\tau\sb{ovr}$ goes to infinity, allowing for wave steepening and breaking effects, which requires finite shock creation time $\tau\sb{sh}$.    In the other words, the main assumption of the WWT theory, roughly formulated as  $\tau\sb{ovr} \ll  \tau\sb{sh}$,  fails for dispersionless acoustic waves even even at small nonlinearity.

 This has   caused   a long-standing debate whether  WWT theory  is applicable for their description \cite{ZakSag70}, or alternatively, if acoustic waves should be viewed as a random collection of weak shocks leading to the Kadomtsev-Petviashvilli spectrum \cite{KadPet73}. There is a handwaving argument--yet unsupported by rigorous proof--that the  theory  applies to 3D acoustic turbulence because the divergence of wave packets in 3D space plays a role similar to the wave dispersion in preventing wave breaking. It was argued first in \cite{newell_aucoin_1971} and later in \cite{PhysRevE.56.390}, that the WWT   description is indeed possible for 3D weak acoustic turbulence. However, the respective kinetic equation for the spectrum has to be modified so that interactions of non-collinear waves are described correctly. Concerning the 2D case, it is evident that WWT theory is not applicable in its classical form. Indeed, the main equation of the  theory, the wave-kinetic equation, is singular and meaningless in the 2D case. The possibility of an alternative statistical description of weak non-dispersive acoustic 2D sound was claimed in \cite{newell_aucoin_1971}, but it remains an unfinished task. 
 Fortunately, in some important physical applications, 2D sound is regularized by weak dispersion effects, and the use of the classical WWT theory becomes again possible. One of such examples, which we will use in the present Letter, is the acoustic turbulence in 2D Bose-Einstein condensates (BEC). 
 
Recent experiments with 3D BECs have succeeded in creating wave-turbulence states where measurements can be explained using the WWT theory in the fully dispersive limit \cite{Navon_EmergenceTurbulentCascade_2016,Navon_SyntheticDissipationCascade_2019}. On the other hand, much interest has been devoted to studying the dynamics of vortices experimentally in 2D BECs, as such states are close to hydrodynamic 2D classical turbulence \cite{Johnstone_EvolutionLargescaleFlow_2019,Gauthier_GiantVortexClusters_2019}. Unfortunately, experiments of acoustic 2D BECs still lack and no theoretical predictions are available.

 In this Letter, we develop the theory of weak wave turbulence of weakly dispersive 2D sound and obtain a modified wave kinetic equation. We derive a new stationary power-law  flux  spectrum of Kolmogorov-Zakharov  type, which corresponds to a cascade of energy from large to small spatial scales. We then determine the   flux  spectrum exponent and the value of the prefactor constant analytically. This prediction can not be naively guessed by dimensional arguments, unlike the existing 3D results. Our theory is then tested and validated with numerical simulations of weakly nonlinear sound in the 2D forced-dissipated Gross-Pitaevskii (GP) equation  \eq{GPE} which is a dynamical model for BEC.

Let us consider the classical wave-kinetic equation   describing evolution of the wave action spectrum $n_{\B k} =  n({\B k},t) $  (with ${\B k}$ and $t$ being the wavenumber and time)  driven by three-wave resonant interactions \cite{ZLF,Nazarenko_2011},
 \begin{equation}
 \frac {\partial n_{\B k}}{\partial t}= \mbox{St}_{\B k}\ 
 \end{equation}
with the wave-collision integral
  \begin{subequations}\label{eq:KE}
    \begin{eqnarray}  \label{St}
  \mbox{St}_{\B k}&=& 2 \pi \int \left(\mathcal{R}^{\B k}_{\B {12}} -\mathcal{R}^{\B 1}_{\B {k2}} -\mathcal{R}^{\B 2}_{\B {k1}} \right) d \B {k}_1 d \B {k}_2\,, \\  \nonumber
    \mathcal{R}^{\B k}_{\B {12}} &=&  \C N ^{\B k}_{\B {12}}      \delta^{\B {k}}_{\B {12}}   \delta(\omega^{\B {k}}_{\B{12}})\,,~~~~~~~~~~~~\\  \label{Nk12}
    \C N ^{\B k}_{\B {12}}&=&|V^{\B k}_{\B {12}}|^2  \big 
     [ n_{  {\B k}_1} n_{  {\B k}_2} - n_{  {\B k}} n_{  {\B k}_1} - n_{  {\B k}} n_{  {\B k}_2} \big]
     \,, \\ \label{dk-k12}
      \delta^{\B k}_{\B{12}} &=&     \delta( {\B k}  - \B {k}_1 - \B {k}_2)\,, \ \nonumber 
    \delta(\omega^{\B {k}}_{\B{12}})= \delta(\omega_ {\B k} - \omega_{  {\B k}_1} - \omega_{  {\B k}_2}) \,,
    \end{eqnarray}\end{subequations} 
$ \omega_{\B k} =\omega({\B k})$  and $V^{\B k}_{\B {12}} \equiv V({\B k},\B {k_1}, \B {k_2})$  is a three-wave interaction amplitude, that
 depends on the particular type of waves. 
 
 The central object in turbulence is the 1D energy spectrum defined as the distribution of energy in $k=|{\B k}|$, so that the energy (per unit of mass) is defined as
$ E=
 \int E(k) dk$.
It is related to the waveaction spectrum as
 $E(k) = 4\pi \omega_k k^2 n_k$
 in 
 3D
 and $E(k)= 2\pi \omega_k k n_k$
 in 2D.
 
 In this Letter, we consider turbulence of weakly dispersive acoustic waves
 for which
 \begin{equation}\label{DL}
\omega_{k}   = \c \, k [1+ (a\,k)^2]\ .
\end{equation}
Here  $a=$~const is a dispersion length such that $a\, k\ll 1$. The interaction amplitude $ V^{k}_{ {12}}$ in case acoustic hydrodynamics \cite{ZakSag70,Dyachenko:1992aa,Zakharov:2005aa,Nazarenko_2011} is given by
 	\begin{equation}\label{V}
 V^{k}_{ {12}}=V_0 \sqrt { k  k_1  k_2}, \quad \quad 
 V_0=\hbox{const}.
 	\end{equation}
 
 To find the energy spectra, note that    \Eq{eq:KE}  conserves the total energy of the system and, therefore, it can be rewritten as the following  continuity equation  for $E(k)$,
 \begin{equation}\label{CEa}
 \frac{\partial E(k)}{\partial t} + \frac{\partial \varepsilon_k}{\partial k}=0\,,\,{\rm where}\,\,  \varepsilon_k= -  \int_{k_1<k} \mbox{St}_{\B k_1 } \omega_{k_1}\, d{\B k}_1.
 \end{equation} 
  Dimensionally, the energy flux $\varepsilon(k)\propto St_k \propto n_k^2$, so that $n_k$ and $E(k)  \propto \sqrt{\varepsilon}$. Assuming full self-similarity (i.e. no other dimensional parameters enter into the game), and that $\varepsilon_k$ is independent of $k$ in an inertial range of scales,  one can reconstruct $E(k)$ from the dimensional reasoning up to a dimensionless constant $C$,
 \begin{subequations}\label{3w} \begin{equation}\label{3g}
  E(k)= C\sqrt{\varepsilon \omega_k}/ k^2\,
  \ .
 \end{equation}
For example, for non-dispersive 3D acoustic waves with $\omega_k=\c k$ we recover the Zakharov-Sagdeev spectrum \cite{ZakSag70},
 \begin{equation}\label{SZ}
 E(k)\propto k^{-3/2}\,,
 \end{equation}\end{subequations}
 which is a      flux  spectrum   describing an energy cascade from large to small scales. It can also be obtained as an exact solution of 
 \Eq{eq:KE}  \cite{ZakSag70}.
 
 Recall that   non-dispersive acoustic waves admit triad wave number and frequency resonances for colinear wavevectors only, i.e. $\B k \| \B k_1 \| \B k_2$. In this case the arguments of $   \delta^{\B k}_{\B{12}}$ and $\delta(\omega^k_{12})$ in \Eq{eq:KE} for St$_{\B k}$ coincide which creates a singularity. Fortunately, in 3D, this singularity is integrable and St$_{\B k}$ remains finite. Therefore  spectrum \Eq{SZ} 
  in 3D is a valid solution (see \cite{newell_aucoin_1971,PhysRevE.56.390,Zakharov:2005aa} for further discussions). 
  In 2D, the singularity is NOT integrable and  
  \Eq{eq:KE} for the non-dispersive case is therefore invalid. Furthermore, the energy spectrum \eqref{SZ} obtained dimensionally is wrong in 2D  because the extra dimensional parameter $a$  becomes essential (unlike  in 3D).

  To solve the problem of acoustic turbulence in 2D, we take into account the dispersion correction in the frequency \eqref{DL}. We choose a reference system with $\B k \| \widehat {\B x} $.   Integrating in Eq.~\eqref{eq:KE} $\mathcal{R}^{\B k}_{\B {12}} $ over $\B k_2$ with the help of $ \B   \delta^{\B k}_{\B{12}}$ and over $ k_{1y}$ using $ \delta(\omega^k_{12})$.  We get
   \begin{subequations}\label{D}
\begin{align}\begin{split}\label{eq:deltas}
& \Delta ^{\B k}_{\B {12}}      \ADD{\defeq} \int_{\B k_2, 
k_{1y}} \delta(\omega^k_{12})    \delta^{\B {k}}_{\B {12}}   d \B k_1d\B k_2 \\ 
& = |\partial_{k_{1y}}(\omega_\B {k} - \omega_\B {k_1} - \omega_\B {k - k_1})|^{-1} dk_{1x} 
 \approx  \frac{k_1k_2 \,dk_{1x}}{ c \sb s \,k \, |k_{1y}|} \,, 
\end{split}\end{align}
where we retained the leading order in $ak\ll 1$ only. Analyzing resonance conditions \eqref{CL} with the the dispersion law\,\eqref{DL}, for small $ak$ we find $k_{1y}\approx \sqrt{6} a k_1 k_2$ (see Supplemental Material). Substituting this  into \Eq{eq:deltas}, we have
\begin{equation}\label{Db}
\Delta ^{\B k}_{\B {12}}\approx  d k_{1x}\big / (\sqrt 6 \, c\sb s\, ak)\ .
\end{equation}
\end{subequations}
We can use this expression to modify the dimensional analysis, which now involves an extra dimensional quantity $a$.
For this, it is only important to observe that $\Delta ^{\B k}_{\B {12}}\propto 1/a$.  
 Therefore $\hbox{St}_{ \B k} \propto  n_k^2 / a$  which, together with St$_{ \B k}\propto \varepsilon$ (arising from \Eq{CEa}),  gives  $n_k\propto \sqrt {\varepsilon \, a}$. This leads to the following energy spectrum of weakly dispersive 2D acoustic turbulence,
\begin{equation}\label{GNL}
E(k)= C_1 \sqrt{\varepsilon \, a\, \omega_k}\big /  k^{3/2}= C_1 \sqrt{\varepsilon\, a\, \c}\big /  k \,,
\end{equation}
where $C_1$ is a dimensionless constant.
Note that the spectrum $E(k)\propto 1/k$  was suggested in \cite{Dyachenko:1992aa} without proving that it is a solution of the kinetic equation.  Below, we will prove that \eqref{GNL} is the unique stationary power-law  flux   solution of  \Eq{eq:KE}    and  find $C_1$ analytically.
%
 To do this, we compute   $\Delta ^{\B 1}_{\B {k2}}$ and $\Delta ^{\B 2}_{\B {k1}}$, similarly to \Eqs{eq:deltas} (see the  Eqs.\,(27) in  Supplemental Material) and present the collision term in \,\eqref{eq:KE} in simpler form,
  	\begin{align}\begin{split} \label{KE}
 	\mbox{St}_{\B k} = & \frac {2\pi} {\sqrt 6\, c\sb s \, ak}\int d {k_1}d   {k_2} \Big[ \mathcal{N}^{\B k}_{\B {12}}\delta(k-k_1-k_2)\\ 
  -&\mathcal{N}^{\B 1}_{\B {k2}}\delta(k_1-k-k_2) -\mathcal{N}^{\B 2}_{\B {k1}}\delta(k_2-k -k_1)\Big ] \ .
   	\end{split}\end{align}
 	
%
  
 Let us seek a stationary solution of    \Eq{eq:KE}  in form 
 	\begin{equation}\label{nx}
  n_k =A  k^{-x}, \quad x=\hbox{const}\,.
 \end{equation} 
 There is, of course, a trivial solution with 
 $x=1$  corresponding to the thermodynamic
 energy equipartition, but we will be interested in non-zero flux states only.
  Assuming that the integrals in \Eq{KE} converge (which is to be checked {\em a posteriori}), let us apply the Kraichnan-Zakharov transform 
 \begin{equation}\label{eq:ZT_2}
 k_1 = \frac{k^2}{\tilde{k}_1}  \,, \;
 k_2 = \frac{k \tilde{k}_2}{\tilde{k}_1}\, \;\; \hbox{and}
 \;\;
 k_2 = \frac{k^2}{\tilde{k}_1}  \,, \quad
 k_1 = \frac{k \tilde{k}_2}{\tilde{k}_1}
 \end{equation}
 to the second and the third terms in \Eq{KE} respectively.
 This gives (after dropping tildes on the integration variables for uniformity of notations): 
 \begin{align} \begin{split} \label{St1}
\mbox{St}_{\B k}  = &  \frac {2\pi} {\sqrt 6\, c \sb s \, ak}\int d k_1 dk_2 \,\mathcal{N}^{\B k}_{\B {12}}\\  
 & \times \Big[ 1- \Big (\frac{k_1}{k}\Big)^{-y} - \Big (\frac{k_2}{k}\Big)^{-y}\Big ]\delta(k-k_1-k_2)\,, 
 \end{split}	\end{align}
 with $y=5-2x$.  Thus,  $\mbox{St}_{\B k}=0$ if $y=-1$, so that  $x=3$ and thus 
 \begin{equation}\label{13}
  n_k\propto \frac 1{k^{3}},\,\hbox{so} \quad E(k)= 2\pi k \omega_k n_k {\approx  2\pi k^2 c_s n_k}\propto \frac 1k \ .
 \end{equation} 
This is an exact stationary solution of the   \Eqs{eq:KE}  and \eqref{KE}  coinciding  with the dimensional prediction \eqref{GNL}.    

To demonstrate  existence and uniqueness of  solution\,\eqref{13}, let us non-dimensionalize the
collision integral for arbitrary $x$: 
\begin{subequations}\begin{align}
\mbox{St}_k {}=& \frac {2\pi A^2 V_0^2}{\sqrt 6 \, a c_s} I(x) k^{3-2x}\,,\\
\begin{split}
I(x){}=& \int_0^1 \!\! q (1-q)    (q^{-x}(1-q)^{-x}-q^{-x}-(1-q)^{-x} )\\  
&  \times 
( 1-q^{-y} -(1-q)^{-y} ) dq.
\end{split}\end{align}\end{subequations}
This integral converges if and only if $2 < x < 4$ or $x=1$, as we prove in the  Supplemental Material.  In \Fig{f:2} we show the plot $I(x)$ obtained numerically for the convergence range $2 < x < 4$. As we see, $x=3$ is the only point at which $I(x)=0$, which proves the uniqueness of the stationary power-law   flux solution
\eqref{GNL}. The existence of the solution amounts to the finiteness of the constant $C_1$.
 We compute $C_1$ in Sec. III.C of the Supplemental Material by substituting  \eqref{St1}
	into the definition of the flux \eqref{CEa}.
	The result is as follows
	\begin{equation}\label{res}
		\varepsilon_k= \frac{4 \pi^4 A^2 V_0^2}{\sqrt{6}  \,a }\,, \quad C_1 = \frac{ 6^{1/4} \, \sqrt{ c_s} }{\pi V_0}\  
	\end{equation}  
	\begin{figure}
	\center{
		\includegraphics[width=.7 \linewidth]{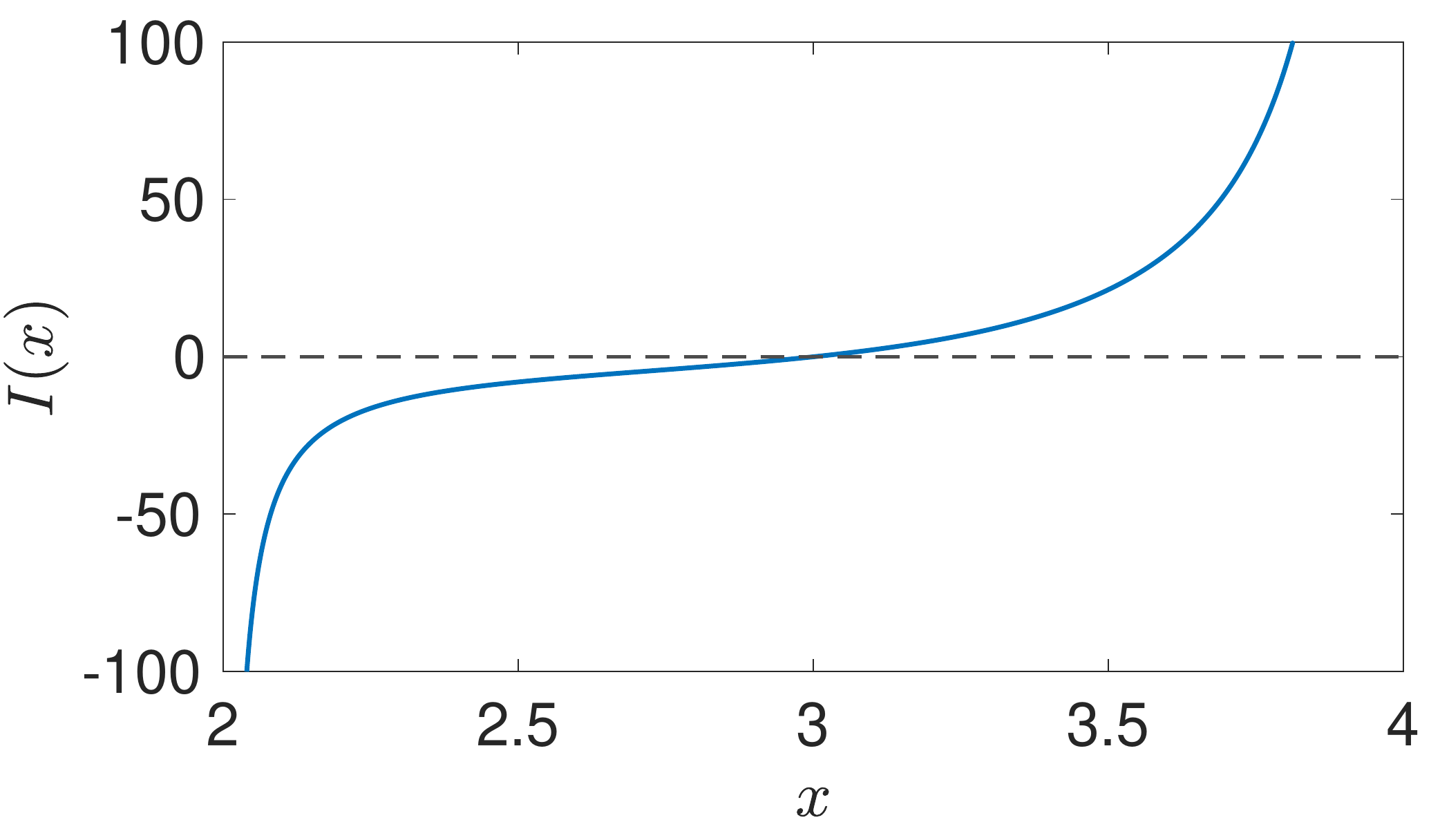}
		\caption{\label{f:2}}Integral \eqref{St1} in the window of convergence.}
\end{figure}
Notice that   \Eq{eq:KE}  is valid for sufficiently small  nonlinearities  and  stochasticity of the phases\,\cite{Nazarenko_2011,ZLF}.    
  Let us define the interaction frequency 
  $\gamma_k$
  as a frequency with which the wave packets are destroyed by the nonlinear interactions. It will also correspond to the nonlinear frequency broadening: a characteristic width of the time-Fourier spectrum at a fixed $\B k$. 
	Applicability of    \Eq{eq:KE}   requires  frequency $\gamma_k$ to be smaller than the characteristic frequency of interacting waves $\delta \omega_k$. For the 3D waves,  we roughly  (and definitely  not rigorously) may take  $\delta \omega_k =
	  \omega_k$. However, due to the non-integrable singularity of the non-dispersive 2D system, in 2D we should take only the dispersive part of the frequency, 
  $	\delta \omega_k =  c\sb s a^2 k^3  $,  ignoring its linear part $c\sb s k$, disappearing in the reference system, co-moving with velocity $c\sb s$ in the $\B k$-direction. 
 On the other hand, for the wave turbulence to be considered acoustic, the dispersion must remain a small correction, i.e. $a  k \ll 1$.
   
  To test our theoretical predictions, we perform direct numerical simulations of the 2D    GP  equation     for the complex wave function $\psi$. Written in terms of the healing length $\xi$, the speed of sound $c_s$ and the bulk density $\rho_0$,   this equation  reads
  \begin{equation}
    i \frac{\partial\psi  }{\partial t}=\frac{c_s}{\sqrt{2}\xi}\Big [ -\xi^2\nabla^{2}  +\frac{1}{\rho_0}|\psi|^{2} - 1\Big ]\psi +i D_c\nabla^{6}\psi+F({\bm r},t) ,
    \label{GPE}
    \end{equation}
  where we have also included a large-scale forcing $F$ and a hyper-viscous dissipation term. The healing length and the speed of sound both depend  on the physical properties of the superfluid and on $\rho_0$, and they can be chosen arbitrarily in the dimensionless GP equation. 

The GP equation  is a well established model for BEC and it can be mapped to and effective compressible irrotational fluid flow via the  Madelung transformation,\\ 
$~~~~~~~~~~~~~~\psi(\B r,t) = \sqrt{\rho(\B r,t)} \exp[i \phi(\B r,t)/\sqrt{2}\c\xi]$\,, \\ with $\rho(\B r,t)$ and $\phi(\B r,t)$ being the fluid density and the velocity potential respectively. Perturbations about a still fluid with uniform density $\rho(\B r,t) \equiv \rho_0 $  behave as a dispersive sound with frequency given by the Bogoliubov dispersion relation, $\omega_k = \c k \sqrt{1+  \xi^2 k^2 /2}  $. 
In the weakly dispersive limit $\xi k \ll 1$, it becomes $\omega_k = \c k (1+  \xi^2 k^2 / 4) $ i.e. the dispersion relation (\ref{DL}) with $a =\xi/2$. 
In this limit, the three-wave interaction coefficient of \Eq{GPE} 
 is of the form \eqref{V}, although some ambiguities and discrepancies in the value of the coefficient $V_0$ can be found in the previous works~\cite{ZakSag70,Dyachenko:1992aa,Zakharov:2005aa,Nazarenko_2011}. In   Sec.I.B of the  Supplemental Material, we provide the corrected derivation which leads to  $V_0 = 3\sqrt{\c}/4\sqrt{2}$.
 Substituting this into Eq.~\eqref{res} we have the following prediction for the pre-factor constant,
 \begin{equation}
     C_1= \frac{2^{11/4}}{3^{3/4} \pi}{\approx 0.94}\ \label{Eq:valC1}.
 \end{equation}
 
We simulate \Eq{GPE} using the standard pseudo-spectral code FROST \cite{mullerIntermittencyVelocityCirculation2021} in a periodic domain of size $L$ using $N_c^2=1024^2$ and $N_c^2=512^2$ collocation points, denoted by Run 1 and Run 2 respectively. The nonlinear term is de-aliased twice with $2/3-$rule following the scheme introduced in \cite{krstulovicEnergyCascadeSmallscale2011} in order to  conserve momentum (in addition to the energy and the number of particles) in the ideal case (with $F=D_c=0)$. 
\ADD{The Fourier transform of the forcing $F$ obeys the Ornstein–Uhlenbeck process $\mathrm{d}F_{\bf k}=-\alpha F_{\bf k} \mathrm{d}t+f_0\mathrm{d}W_{\bf k}$}, where $W_{\bf k}$ is the Wiener process. The forcing acts only on wavenumbers such that $2\pi\le k L\le 3\times 2\pi$. In addition, the condensate amplitude is kept constant during the evolution.
We set the initial data with uniform condensate with $|\psi_0|^2 = \rho_0$, the forcing then adds the acoustic disturbances, and we evolve the system until it reaches a steady-state. We then perform averages over time. 
In numerics, we have set $\c=1$, $\rho_0=1$ and $\xi=2L/N_c$. 
For forcing and dissipation we set $\alpha=1$, $f_0=1.25\times10^{-4},5\times10^{-4}$ and $D_c=4.1\times10^{-15}, 2.1\times10^{-11}$ for Run 1 and 2 respectively.

In absence of forcing and dissipation,  \ADD{\Eq{GPE}} conserves the total energy (Hamiltonian) of the system. The energy per unit of mass, written in terms of the hydrodynamic variables, consists of the kinetic, internal and quantum energies~\cite{Nore97}:
\begin{equation}
  E =\frac{1}{L^2\rho_0}  \int \left[\frac{\rho}{2}  (\nabla \phi)^2 + \frac{c^2}{2\rho_0} (  \rho-\rho_0)^2  + c^2\xi^2( \nabla  \rho)^2 \right] \mathrm{d}^2 \B r .
 \label{H}
\end{equation}
In the dispersiveless limit ($\xi\to0$), we retrieve the standard energy for a for a compressible, isentropic, irrotational fluid \cite{Landau1987Fluid}. The total energy spectrum is computed writing the energy as usual in quantum turbulence~\cite{Nore97}. 
We also calculate the $k$-space energy flux $\epsilon(k)$ directly using    \Eq{GPE}  (see the  Eq.(31) of  Supplemental Material  for exact definitions).

In Fig.~\ref{f:3} we show the fluxes and spectra for Run 1 and Run 2. 
\begin{figure}
		\includegraphics[width=.99\linewidth]{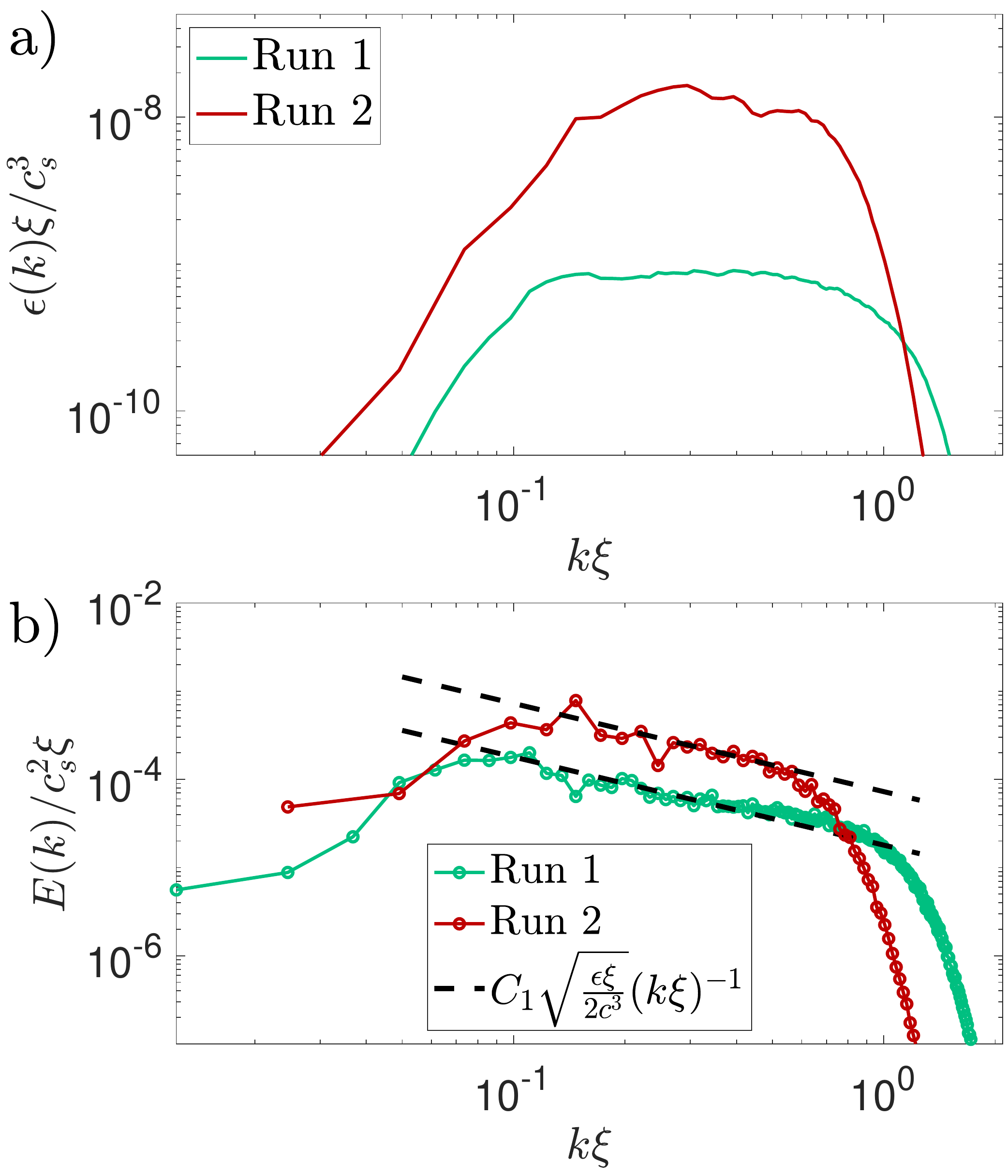}
		\caption{\label{f:3}} {\bf a)} Energy flux. {\bf b)} Energy spectra. Dashed lines correspond to theoretical prediction \eqref{GNL} and \eqref{Eq:valC1} using the corresponding flux values in the inertial range. 
\end{figure}
We see that the fluxes have a pronounced plateau which indicates the presence of an inertial range (free of forcing and dissipation effects). Both runs display a stationary power-law spectrum. Remarkably, both, the power-law exponent and the pre-factor $C_1$ (calculated based on the averaged flux in the inertial range), closely agree with the theoretical predictions \eqref{GNL} and \eqref{res}. 

\begin{figure}
		\includegraphics[width=.99\linewidth]{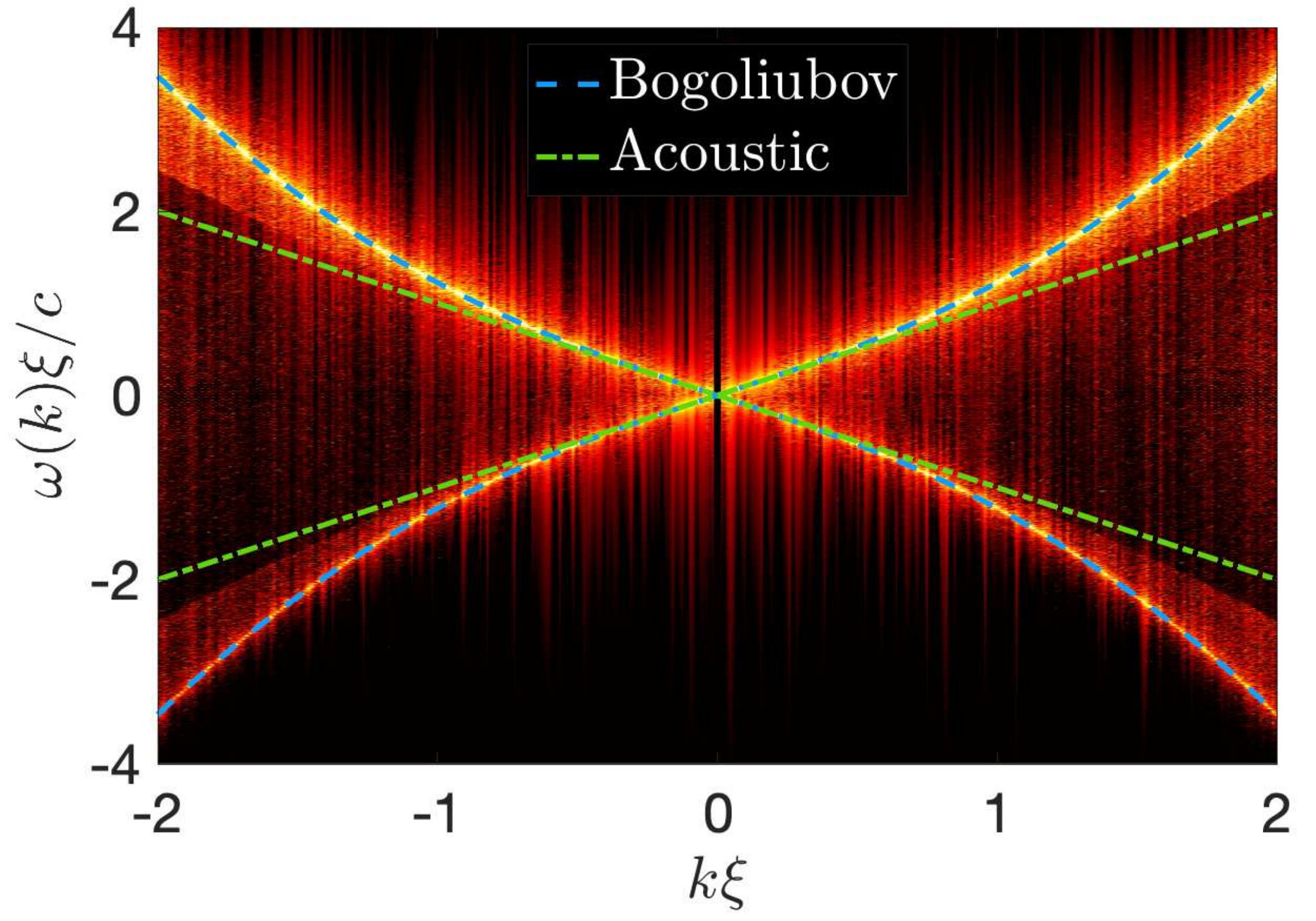}
		\caption{\label{f:5} Spatio-temporal spectrum $S(\omega,k)=|\hat{\psi}(\omega,k)|^2$ normalized by the time-averaged spectra $|\hat{\psi}(t,k)|^2$ of Run1 ($1024^2$).
    The dashed and dot-dashed lines show the Bogoliubov and the pure acoustic dispersion relation. }
\end{figure}

\begin{figure}
	\center{
		\includegraphics[width=.99\linewidth]{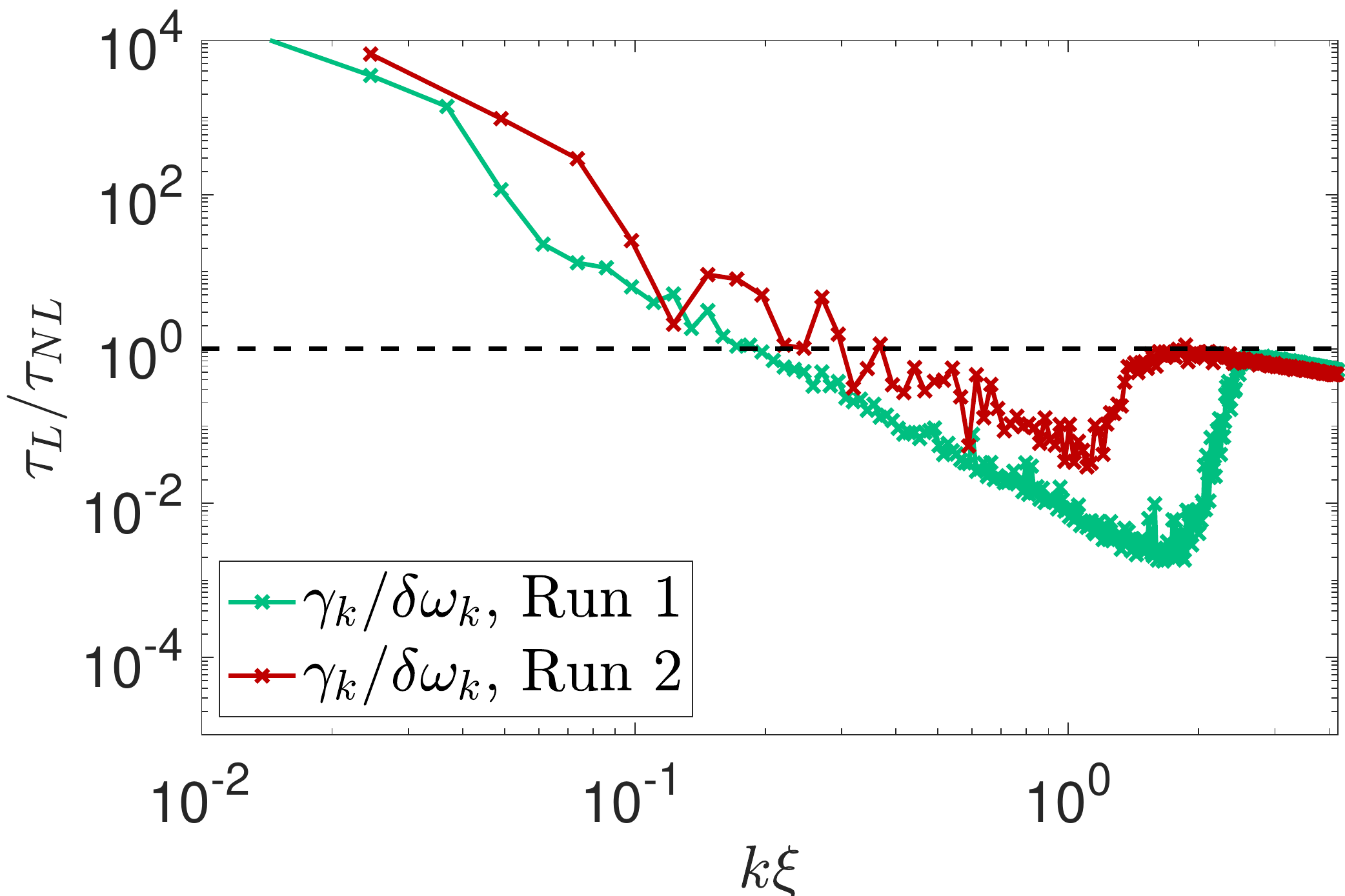}
		\caption{\label{f:4}} Linear to nonlinear time ratios for Run1 ($1024^2$) and Run2 ($512^3$) computed from the STS in Fig.~\ref{f:5}.}
\end{figure}

In Fig.~\ref{f:5} we show the spatio-temporal spectrum  for Run1. 
We see that  this spectrum follows closely the Bogoliubov dispersion law, which indicates that the nonlinearity is sufficiently weak.  The  $\omega$-width of  the spectrum  at each fixed $k$ represents the nonlinear frequency broadening; we define it as\\
$\gamma_k=\Big[\int_0^\infty (\omega-\omega_k)^2 |\hat{\psi}(\omega,k)|^2\mathrm{d}\omega/\int_0^\infty |\hat{\psi}(\omega,k)|^2\mathrm{d}\omega\Big]^{1/2}$.
In Fig.~\ref{f:4} we show the ratios $\gamma_k/\delta \omega_k$, where $\delta \omega = \c a^2 k^3$ is the dispersive correction.
Recall that the WT theory is applicable when $\gamma_k/\delta \omega_k \ll 1$.
We see that this quantity is indeed small in the scaling range in Run1, and
only marginally small in a rather narrow range in Run2. This indicates that the WT theory has a good predictive power even when the formal applicability condition is on the borderline of validity.

 The main result of our paper is the 1D energy spectrum of 2D weakly dispersive acoustic waves, \Eq{GNL}, found as the unique stationary constant-flux solution of the  kinetic equation  \eqref{eq:KE} with convergent collision integral. 
From the physical view point such a convergence means that the main contribution to the energy balance of waves with wavenumber $k$ comes from their energy exchange with the ``neighbouring" waves with wavenumbers $k'$ of the order of $k$.  In the language of hydrodynamic turbulence we are dealing here with  the  step-by-step cascade energy transfer,   local in the wavenumber space. 
 We found the energy flux to be positive, meaning that the energy is transferred from small to large $k$ i.e. it is a \textit{direct energy cascade}.

We tested our analytical predictions by  numerical simulations  of the forced-dissipated  GP equation  \eqref{GPE} in the presence of a strong condensate. The analytically found  spectrum \eqref{GNL} was confirmed by the numerics, including both the power-law exponent and the pre-factor $C_1$ without any adjustable parameter. Such a double validation is a rare success in the  theory  of wave turbulence, where numerical tests were attempted by  numerical simulations for various types of waves but, in most cases, only the spectrum exponent was confirmed. Wave turbulence is therefore a valid and productive approach for describing \ADD{2D} superfluid BEC turbulence where interacting sound waves represent the principal mechanism of energy dissipation. \ADD{Since measurements of the spectrum are experimentally accessible in BEC \cite{Navon_EmergenceTurbulentCascade_2016,Navon_SyntheticDissipationCascade_2019}, our results present verifiable predictions which could guide future experiments.} 
The focus of the present Letter was on the weak  turbulence  of 2D acoustic waves, and  the strong turbulence regimes would be an interesting subject for future studies.


\begin{acknowledgments}
  This work was supported by the Agence Nationale de la Recherche through the project GIANTE ANR-18-CE30-0020-01, by the Simons Foundation Collaboration grant Wave Turbulence (Award ID 651471) 
  and by NSF-BSF grant \# 2020765.
  This work was granted access to the HPC resources of CINES, IDRIS and TGCC under the allocation 2019-A0072A11003 made by GENCI.
  Computations were also carried out at the Mésocentre SIGAMM hosted at the Observatoire de la Côte d'Azur.
\end{acknowledgments}


\clearpage
\appendix


\begin{widetext}
  \begin{center}
    \Large
    \textbf{Energy spectrum of two-dimensional acoustic turbulence: Supplemental Material}
  \end{center}

  \renewcommand\thefigure{S\arabic{figure}}    
  \setcounter{figure}{0}

\section{Hamiltonian formulation of acoustic turbulence}

In this section we review the Hamiltonian formulation of acoustic turbulence and obtain the interaction term represented by  $V^{k}_{ {12}}$ in Eq.(5) of the main text. In the acoustic limit, this term is well known to be $V^{k}_{ {12}}=V_0\sqrt{k k_1 k_2}$~\cite{ZakSag70,Dyachenko:1992aa,Zakharov:2005aa,Nazarenko_2011}, however  the constant $V_0$ in these references takes different values. Here, we will carefully derive  its value.

The starting point is the action  (per unit of mass) for a compressible, isentropic, irrotational fluid \cite{Landau1987Fluid} that reads:
\begin{align}
  \mathcal{S} = \frac{1}{\rho_0 L^2}\int dt d^2x \left[ -\rho \dot{\phi}  - \frac{\rho}{2} (\nabla \phi)^2 - \frac{\c^2}{2\rho_0}(\rho - \rho_0)^2 \right], \label{Eq:LAG}
\end{align}
where $\rho_0$ is the bulk density and $c_s$, as it will be clear later, is the speed of sound. Note that the dimensions of the fields are $[\phi]=L^2/T$ and $[\rho]=M/L^2$.

Varying the action with respect to $\rho$ and $\phi$ and we obtain the fluid equations,
\begin{eqnarray}
  \dot{\rho}+\nabla\left(\rho\nabla\phi\right)&=&0,\\
  \dot{\phi}+\frac{1}{2}\nabla\phi^2&=& -\frac{\c^2}{\rho_0}(\rho-\rho_0). 
\end{eqnarray}
Acoustic waves are readily obtained by linearizing the equations about $\phi=0$ and $\rho=\rho_0$, which leads to the wave equation $\ddot{\phi}=\c^2\nabla^2\phi$. 

Note that the action \eqref{Eq:LAG} is not written in a Hamiltonian way. Making the following change of variables $\rho = \rho_0A^2$ and $p=2 A \phi$, after substituting in \eqref{Eq:LAG}, we obtain 
 \begin{eqnarray}
  \mathcal{S} = \int dt \frac{d^2x}{L^2}\frac12\left(A \dot{p}-\dot{A}p\right) - \int dt H\quad \textrm{with} \quad   H = \int \frac{d^2x}{L^2} \left[\frac18\left( \nabla p - \frac{p\nabla A}{A}\right)^2 + \frac{\c^2}{2}(A^2 - 1)^2 \right],\label{Eq:LAG_pA}
  \end{eqnarray}
  where the equation of motion are now given by
  \begin{eqnarray}
  \dot{p} =\frac{\delta H}{\delta A} &,\quad& 
  \dot{A} =-\frac{\delta H}{\delta p}.
  \end{eqnarray}
We remark that the units of the new fields are $[A]=1$ and $[p]=L^2/T$. The Hamiltonian per unit of mass has units $[H]=L^2/T^2$ as usual in hydrodynamics.
   
  \subsection{Acoustic waves}
 Waves are obtained by making $A\to 1+\tilde{A}$ and $p\to \tilde{p}$. Dropping tildes and keeping the terms up to the cubic order,
we rewrite the Hamiltonian as $H=H_2+H_3$, where the second and third order terms are
 \begin{eqnarray}
    H_2  = \int \frac{d^2 x}{L^2} \left[ \frac{1}{8} (\nabla p)^2 + 2\c^2 A^2 \right]
    &,\quad&
    H_3  = \int \frac{d^2x}{L^2} \left[2\c^2A^3 - \frac{1}{4} p(\nabla p)\cdot (\nabla A) \right].
 \end{eqnarray}
We now assume that the fields are periodic and write them as $p({\bf x})=\sum_{\bf k} p_{\bf k}e^{i {\bf k}\cdot{\bf x}}$ and $A({\bf x})=\sum_{\bf k} A_{\bf k}e^{i {\bf k}\cdot{\bf x}}$. The Hamiltonian and the action become:
\begin{eqnarray}
  \mathcal{S} &=& \int dt \sum_\k\frac12\left(A_k \dot{p}_\k^*-\dot{A}_\k p_\k^*\right) - \int dt (H_2+H_3) , \\
  H_2 & =& \sum_\k \frac{1}{8}k^2 |p_\k|^2 + 2\c^2 |A_\k|^2, \\
  H_3 & =& \sum_{1,2,3} 2 \c^2A_1 A_2 A_3 \delta_{1,2,3} + \frac{1}{4}p_1 p_2 A_3 {\bf k}_2 \cdot {\bf k}_3 \delta_{1,2,3},
  \label{Eq:ActionWavesFourier}
  \end{eqnarray}
  where $\delta_{1,2,3}$ is $1$ if $\k_1+\k_2+\k_3=0$, and $0$ otherwise.
  
  In order to write the Hamiltonian and the action in the canonical form, we perform the following change of variables
  \begin{align}
  p_\k = i \frac{1}{\sqrt{2}} \left(\frac{\alpha}{k^2}\right)^\frac14\left(a_\k - a^*_{-\k}\right), \\
  A_\k =  \frac{1}{\sqrt{2}} \left(\frac{k^2}{\alpha}\right)^\frac14\left(a_\k + a^*_{-\k}\right),
  \end{align}
  where $\alpha=16\c^2$. The value of this coefficient is set in order to kill the off-diagonal terms in $H_2$. 
At the leading order, the action becomes
  \begin{align}
  \mathcal{S}_2 & = \int dt \frac{i}{2}\sum_\k \left(\dot{a}_\k a^*_\k - a_\k \dot{a}_\k^* \right) -\int dt H_2\,,\quad\textrm{with }H_2=\sum_\k \c k|a_\k|^2=\sum_\k \omega_k|a_\k|^2.
  \end{align}
  Then,
  \begin{align}
  \frac{\delta \mathcal{S}_2 }{\delta a^*}=0 \implies i \dot{a}_\k = \frac{\partial H}{\partial a^*_\k} = \omega_k a_\k.
  \end{align}
  
  \subsection{$H_3$ terms}
  The cubic part of the Hamiltonian requires some tedious work. Keeping only resonant terms we obtain the following contributions
  \begin{equation}
    \sum A_1 A_2 A_3 \delta_{1,2,3} = \frac{1}{2^{3/2}} \sum_{1,2,3} \frac{\sqrt{k_1 k_2 k_3}}{\alpha^\frac34}(a_1 + a^*_{-1})(a_2 + a^*_{-2})(a_3 + a_{-3}^*)  =  \frac{3}{2^{3/2}}\sum\frac{\sqrt{k_1 k_2 k_3}}{\alpha^{\frac34}}\left(a_1 a_2^* a_3^* + a_1^* a_2 a_3 \right) \delta^1_{2,3},
    \end{equation}
  where $\delta^1_{2,3}=\delta_{-1,2,3}$. The second term requires more manipulations
  \begin{align}
  \sum p_1 p_2 A_3 {\bf k}_2\cdot {\bf k}_3\delta_{1,2,3} & =-  \frac{1}{2^{3/2}}\sum_{1,2,3} \frac{\alpha^\frac14}{\sqrt{k_1 k_2 k_3}}k_3 ({\bf k}_2 \cdot {\bf k}_3)(a_1 - a^*_{-1})(a_2 - a^*_{-2})(a_3 + a_{-3}^*)\delta_{1,2,3} \\
  & =- \frac{1}{2\times2^{3/2}} \sum_{1,2,3} \frac{\alpha^\frac14}{\sqrt{k_1 k_2 k_3}}k_3 ( {\bf k}_1 \cdot {\bf k}_3 + {\bf k}_2 \cdot {\bf k}_3)(a_1 - a^*_{-1})(a_2 - a^*_{-2})(a_3 + a_{-3}^*)\delta_{1,2,3}\\
  & =\frac{1}{2\times2^{3/2}}  \sum_{1,2,3} \frac{\alpha^\frac14}{\sqrt{k_1 k_2 k_3}}k_3^3 (a_1 - a^*_{-1})(a_2 - a^*_{-2})(a_3 + a_{-3}^*)\delta_{1,2,3},
  \end{align}
  where form the second to third line we used the resonant condition $\k_1+\k_2=-\k_3$. Again, keeping only resonant terms, changing summation variables and using symmetries, we can replace inside the sum 
\begin{equation}
  k_3^3 (a_1 - a^*_{-1})(a_2 - a^*_{-2})(a_3 + a_{-3}^*)\delta_{1,2,3}
  \to \left(a_1 a_2^* a_3^*+a_1^* a_2 a_3\right) (k_1^3-2k_3^3)\delta^1_{2,3}
  \to \left(a_1 a_2^* a_3^*+a_1^* a_2 a_3\right) (k_1^3-k_2^3-k_3^3) \delta^1_{2,3}
\end{equation}
Finally, using the resonant condition $k_1=k_2+k_3$, we have   $k_1^3-k_2^3-k_3^3=(k_2+k_3)^3-k_2^3-k_3^3=3(k_2+k_3)k_2k_3=3k_1k_2k_3$. Gathering all the terms
  \begin{equation}
    H_3  = \frac{1}{2^{3/2}} \sum (a_1a_2^* a_3^* + c.c)\delta^1_{2,3} \sqrt{k_1 k_2 k_3} \left[2\c^2\frac{3}{\alpha^{3/4}} +\frac{1}{4}\frac{3\alpha^{\frac14}}{2} \right] =\sum V^1_{23}(a_1a_2^* a_3^* + c.c)\delta^1_{2,3},
  \end{equation}
  where $ V^1_{23}= V_0\sqrt{k_1 k_2 k_3}$, with 
  \begin{align}
   V_0=\frac{1}{2^{3/2}}\left[2\c^2\frac{3}{\alpha^{3/4}} +\frac{1}{4}\frac{3\alpha^{\frac14}}{2} \right]= \frac{3 }{4\sqrt{2}}\sqrt{\c},\label{Eq:V0}
  \end{align}
  the formula used to obtained Eq.~(18) from Eq.~(16) in the main text.

\section{\label{A1} Analysis of the collision term}
 
\subsection{Analysis of  $\C R^{  k}_{12}$ term} 

\begin{figure}
	\center{
 	\includegraphics[width=.3\linewidth]{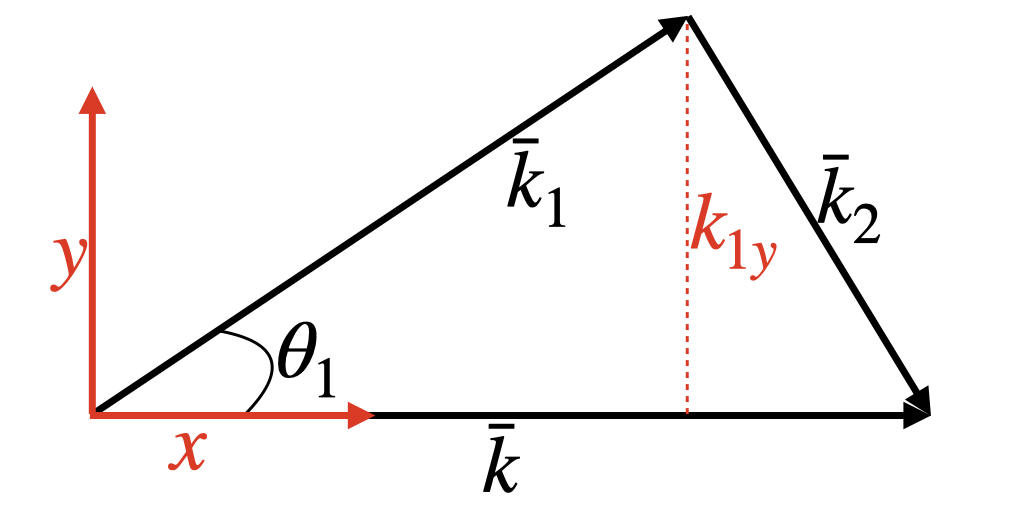}
		\caption{{\label{f:1}}}  { Wave vector triad. We choose a coordinate system such that $\bar{k}$ is aligned with the $x$-axis. $\theta_1$ is the angle between $\bar{k}$ and $\bar{k}_1$,  $k_{1y}$ is the $y$-component of $\bar{k}_1$.}}
\end{figure}

According to Eqs.~(3) of the main text,   the first term in the collision integral ($\C R^{  k}_{12}$) contains
\begin{eqnarray}\label{D1}
\Delta ^{\B k}_{\B {12}} \equiv \int_{\B k_2, 
k_{1y}} \delta(\omega^k_{12})  \B  \delta^{\B {k}}_{\B {12}}   d \B k_1d\B k_2 
 \approx \frac{k_1k_2 \,dk_{1x}}{ c \sb s \,k \, |k_{1y}|} \ .
\end{eqnarray}
From Fig.~1  we can see that to leading order
\begin{align}
k_{1y} = k_1 \sin(\theta_{1k}) \approx k_1 \theta_{1k} . \label{eq:k1simp}
\end{align}
To find $\theta_{1k}$ consider the wave number resonance condition,
\begin{eqnarray}\nonumber
\B k_2= \B k -\B k_1  &\Rightarrow&  
k^2_2= k_1^2+k ^2 - 2 \B k_1\cdot \B k     \nonumber
= k_1^2+k ^2 - 2  k_1  k  \cos \theta_{1k} \nonumber
\approx (k -k_1)^2 +  k   k_1    \theta_{1k}^2 \\
&\Rightarrow&  
k_2\approx (k -k_1)  + \frac { k   k_1  } {2 k_2}\,\theta_{1k}^2\ .\label{eq:a1}
\end{eqnarray}
Next, consider the frequency resonance condition,
\begin{eqnarray}  
k_1 + k_2 - k   = -a^2( k_1^3 + k_2  ^3 - k^3), \end{eqnarray}
and substitute $k_2$ from \Eq{eq:a1} to the LHS and $k_2=(k -k_1)$ to the RHS of this equation. This gives  
 $\dfrac{k k_1  \theta_{k1}^2}{2 k_2}= 3 a^2 k k_1 k_2$ or $\theta_{k1}^2= 6 a^2 k_2^2$ and  
 \begin{equation}
 k_{1y}= \sqrt 6  a \,k_1 k_2\ .
 \end{equation}
 Together with \Eq{D1}, this finally gives
 \begin{equation}\label{D11}
 \Delta ^{\B k}_{\B {12}}\approx \frac{~~d k_{1x}}{\sqrt 6 \, c\sb s\, ak} = \frac{\delta (k-k_1 -k_2)d k_{1} d k_2 }{\sqrt 6 \, c\sb s\, ak} ,
 \end{equation}
also shown in Eq. (8.b) in the main text. Here, we have replaced $d k_{1x}$ by $d k_{1 }$ and inserted $\delta (k-k_1 -k_2)  dk_2=1$ to stress that $k=k_1+k_2$ in the used approximation.

\subsection{ Contribution to $\C R^1_{k2}$  and $\C R^2_{k1}$} 

Similar derivations using the wave number and frequency resonance conditions
lead to
\begin{equation}\label{D2_2}
\Delta ^{\B 1}_{\B {2k}}\approx \frac{~~d k_{1x}}{\sqrt 6 \, c\sb s\, ak}= \frac{\delta (k_1-k -k_2)d k_{1} d k_2 }{\sqrt 6 \, c\sb s\, ak}\quad
{\rm and}
\quad
 \Delta ^{\B 2}_{\B {1k}}\approx \frac{\delta (k_2-k -k_1)d k_{1} d k_2 }{\sqrt 6 \, c\sb s\, ak}\ .
\end{equation}
Together, Eqs.~(\ref{D11}) and (\ref{D2_2}) lead to the collision integral (10) of the main text.

\subsection{\label{A2} Proof of the interaction locality}

Convergence of the integral in Eq. (10) of the main text is referred to as the interaction locality property. First,
we note that this integral is trivially convergent
for $x=1$ because the integrand is identically equal to zero.
This exponent corresponds to the thermodynamic energy equipartition state, i.e. a trivial zero-flux equilibrium which we will not be interested in. Thus, below we will consider the cases with 
{$x\ne 1$}.

\subsubsection{Infrared locality}
Consider first the infra-red (IR) locality, i.e. convergence of the  integral in Eq.~(10) of the main text, in the region $k_1\ll k$. We take into account that for the acoustic turbulence $V_{ {12}}^{ k}\propto \sqrt{k\, k_1 \, k_2}$ an integrate over $k_2$ with the help of the $\delta$-functions. Then the leading term is
\begin{eqnarray}
  \mbox{St}_k&\propto& \int _0 ^{k }  \Big (\C N^k_{k_1,{k - k_1}} - \C N ^{k + k_1}_{k,k_1}  \Big )d k_1
  \propto \int _0 ^{k }  k_1  n_{k_1}\Big [ \big (n_{k-k_1}-n_k \big )  - \big (n_{k }-n_{k+k_1} \big ) \Big ] d k_1 \\ \nonumber 
  &=& \int _0 ^{k }  k_1  n_{k_1}\Big [n_{k+{k_1}}+n_{k-{k_1}}- 2 n_k\Big ] d k_1
  \propto \frac{d^2 n_k}{dk^2} \int _0 ^{k }  k_1  n_{k_1} k_1^2  d k_1\propto \int _0 ^{k }  k_1^3 n_{k_1} d k_1\ .
  \end{eqnarray}
We see that 
this integral converges for any $n_k\propto k^{-x}$  with $x<4$ including $x=3$.

\subsection{Ultraviolet locality}
 
Consider now  the ultraviolet  (UV) locality, i.e. convergence of integral in Eq. (9) of the main text, in the region $k_1\gg k$.
Now the  leading term is
\begin{eqnarray}
  \mbox{St}_k&\propto& \int _k ^{\infty}  \Big (\C N^k_{k_1,{k - k_1}} - \C N ^{k + k_1}_{k,k_1}  \Big ) d k_1
  \propto \int _k ^{\infty}  k_1^2  \Big [ n_k n_{{k_1}-k} -n_{k_1} \big (n_{k}+n_{{k_1}-k}\big ) 
   + n_k n_{k_1} -n_{k+{k_1}}\big (n_{k }+n_{{k_1}} \big ) \Big ] d k_1 \\ \nonumber 
  &\approx& n_k \int _k ^{\infty}   k_1^2  \Big [n_{{k_1}-k}-n_{{k_1}+k} \Big ] d k_1
  \approx -2 k \int _k ^{\infty}  k_1 ^2 \frac{d   n_{k_1}}{dk_1}  d k_1\propto \int _k ^{\infty} \frac {k_1^2   d k_1}{k_1^{x+1}}\ .
  \end{eqnarray}
We see that  in the UV-region  this integral converges for any $x>2$, including $x=3$. 

The overall conclusion is that the collision integral converges for $2< x < 4$ and actual scaling exponent $x=3$ is exactly in the middle of the locality window. This phenomenon is called counterbalanced locality of the collision integral, which quite common property of the kinetic equations.  

 {\section{Energy spectrum, energy flux, and constant $C_1$}
}
 {\subsection{Energy spectrum}}
‡

The total energy (Eq. (19) of the main text) can be rewritten as
\begin{equation}
  E =\frac{1}{L^2\rho_0}  \int \left[\frac{\rho}{2}  (\nabla \phi)^2 + \frac{c^2}{2\rho_0} (  \rho-\rho_0)^2  + c^2\xi^2( \nabla  \rho)^2 \right] \mathrm{d}^2 \B r =\frac{1}{L^2\rho_0}  \int \left[\c^2 \xi^2 |\nabla \psi|^2+ \frac{\c^2}{2\rho_0} (  \rho-\rho_0)^2   \right] \mathrm{d}^2 \B r .
 \label{H}
\end{equation}

The energy spectrum is then computed taking into account that the total energy is the sum of two quadratic quantities and using the definition of the cross spectrum of two fields $f$ and $g$ that is defined in terms of their Fourier transform $\hat{f}$ and $\hat{g}$ as $$E_{f,g}(k)=\frac 1{\Delta_k} \sum_{k-\Delta_k/2<|\k|<k+\Delta_k/2} \hat{f}_{\k}\hat{g}_{\k}^*$$ for some small $\Delta_k$. Note that by the Parseval theorem $\int f({\bf x})g^*({\bf x})d {\bf x}= L^2 \sum_k E_{f,g}(k) \Delta_k \approx \int_k E_{f,g}(k) \, dk $. The total energy spectrum is then computed as $E(k)=E_{\rm kin}(k)+E_{\rm int}(k)$, where $\rho_0 E_{\rm kin}(k)=\c^2\xi^2 E_{\nabla \psi,\nabla \psi}(k)$ and $2\rho_0^2 E_{\rm int}(k)=\c^2E_{\rho-\rho_0,\rho-\rho_0}(k)$.

 {\subsection{Energy flux}}
The energy flux   {can be} computed as usual in hydrodynamics \cite{Frisch1995}, but adapting it to GP dynamics as
\begin{equation}
\varepsilon(k)=-\sum_{p=0}^k \left.\frac{\partial E(p)}{\partial t}\right|_{\rm GP}=-\sum_{p=0}^k \left.\frac{\partial E_{\rm kin}(p)}{\partial t}\right|_{\rm GP}+\left.\frac{\partial E_{\rm int}(p)}{\partial t}\right|_{\rm GP},
\end{equation}
where the label GP means that time derivatives are computed using the GP equation (without forcing and dissipation). Namely, we have
\begin{eqnarray}
  \left.\frac{\partial E_{\rm kin}(p)}{\partial t}\right|_{\rm GP} &=&\frac{2 \c^2\xi^2}{\rho_0} E_{\nabla \psi,\nabla d\psi}(k)\\
  \left.\frac{\partial E_{\rm int}(p)}{\partial t}\right|_{\rm GP} &=& \frac{\c^2}{\rho_0^2}E_{\rho-\rho_0,d\rho}(k)
\end{eqnarray}
where $d\psi=-i\frac{\c}{\sqrt{2}\xi}\Big [ -\xi^2\nabla^{2}  +\frac{1}{\rho_0}|\psi|^{2} - 1\Big ]\psi$ and $d\rho=\psi d\psi^*+d\psi \psi^*$. Note that $\lim_{k\to\infty}\varepsilon(k)=0$ because of the energy conservation of the GP equation.

\subsection{ {Dimensionless prefactor $C_1$}}
To compute $C_1$, we substitute Eq.(15)  
into the definition of the flux (6) (both in the main text), substitute $d{\B k}_1= 2\pi k_1 dk_1$  (since the integrand is a function of the $k_1 =|{\bf k}_1|$ and the polar angle is immediately integrated out)  and integrate with respect to $k_1$. This  leads to 
\begin{equation} 
	\varepsilon_k= - \frac{2\pi^2 A^2 V_0^2}{\sqrt{6}\,a}\,\frac{I(x)}{(3-x)} k^{ 6-2x}\ .
	\label{eq:flux}
	\end{equation}
	For actual value $x=3$ this equation has an uncertainties zero divided by zero, which can be resolved according the 
 the L'Hopital rule:
 \begin{equation}
	\lim_{x\to 3}  \frac{I(x)}{(3-x)} =
	\frac{dI(x)}{dx}\Big |_{x=3}= \int_0^1 \frac{12 \log q_1}{q_1-1}dq_1 = 2\pi^2.
	\nonumber
\end{equation}
 Now \Eq{eq:flux}  with $x =3$ give for the energy flux:
 \begin{equation}\label{res1}
	\varepsilon_k= \frac{4 \pi^4 A^2 V_0^2}{\sqrt{6}  \,a }\ .
\end{equation}
 
Thus, \Eq{res1}, together with Eqs.(9) and (11) of the main text finally give: 
\begin{equation}\label{res2}
	  C_1 = \frac{ 6^{1/4} \, \sqrt{ c_s} }{\pi V_0}\ .
\end{equation}
for the  pre-factor $C_1$ in  Eq. (9) of the main text.

\end{widetext}


\begin{thebibliography}{24}
  \expandafter\ifx\csname natexlab\endcsname\relax\def\natexlab#1{#1}\fi
  \expandafter\ifx\csname bibnamefont\endcsname\relax
    \def\bibnamefont#1{#1}\fi
  \expandafter\ifx\csname bibfnamefont\endcsname\relax
    \def\bibfnamefont#1{#1}\fi
  \expandafter\ifx\csname citenamefont\endcsname\relax
    \def\citenamefont#1{#1}\fi
  \expandafter\ifx\csname url\endcsname\relax
    \def\url#1{\texttt{#1}}\fi
  \expandafter\ifx\csname urlprefix\endcsname\relax\def\urlprefix{URL }\fi
  \providecommand{\bibinfo}[2]{#2}
  \providecommand{\eprint}[2][]{\url{#2}}
  
  \bibitem[{\citenamefont{Zakharov et~al.}(2012)\citenamefont{Zakharov, L'vov,
    and Falkovich}}]{ZLF}
  \bibinfo{author}{\bibfnamefont{V.}~\bibnamefont{Zakharov}},
    \bibinfo{author}{\bibfnamefont{V.}~\bibnamefont{L'vov}}, \bibnamefont{and}
    \bibinfo{author}{\bibfnamefont{G.}~\bibnamefont{Falkovich}},
    \emph{\bibinfo{title}{Kolmogorov Spectra of Turbulence, wave turbulence}}
    (\bibinfo{publisher}{Springer}, \bibinfo{year}{2012}).
  
  \bibitem[{\citenamefont{Nazarenko}(2011)}]{Nazarenko_2011}
  \bibinfo{author}{\bibfnamefont{S.}~\bibnamefont{Nazarenko}},
    \emph{\bibinfo{title}{Wave Turbulence}} (\bibinfo{publisher}{Springer Berlin
    Heidelberg, series: Lecture Notes in Physics}, \bibinfo{year}{2011}).
  
  \bibitem[{\citenamefont{Falcon and
    Mordant}(2022)}]{Falcon_ExperimentsSurfaceGravity_2022}
  \bibinfo{author}{\bibfnamefont{E.}~\bibnamefont{Falcon}} \bibnamefont{and}
    \bibinfo{author}{\bibfnamefont{N.}~\bibnamefont{Mordant}},
    \bibinfo{journal}{Annual Review of Fluid Mechanics}
    \textbf{\bibinfo{volume}{54}}, \bibinfo{pages}{annurev}
    (\bibinfo{year}{2022}), ISSN \bibinfo{issn}{0066-4189, 1545-4479}.
  
  \bibitem[{\citenamefont{Galtier et~al.}(2000)\citenamefont{Galtier, Nazarenko,
    Newell, and Pouquet}}]{Galtier_WeakTurbulenceTheory_2000}
  \bibinfo{author}{\bibfnamefont{S.}~\bibnamefont{Galtier}},
    \bibinfo{author}{\bibfnamefont{S.~V.} \bibnamefont{Nazarenko}},
    \bibinfo{author}{\bibfnamefont{A.~C.} \bibnamefont{Newell}},
    \bibnamefont{and} \bibinfo{author}{\bibfnamefont{A.}~\bibnamefont{Pouquet}},
    \bibinfo{journal}{Journal of Plasma Physics} \textbf{\bibinfo{volume}{63}},
    \bibinfo{pages}{447} (\bibinfo{year}{2000}), ISSN \bibinfo{issn}{0022-3778,
    1469-7807}.
  
  
  
 { \bibitem[{\citenamefont{Zakharov}(1972)}]{Zakharov_CollapseLangmuirWaves_}
  \bibinfo{author}{\bibfnamefont{V.E.}~\bibnamefont{Zakharov}},
    \bibinfo{journal}{Soviet Physics—JETP}
    \textbf{\bibinfo{volume}{35}}, \bibinfo{pages}{908-914} (\bibinfo{year}{1972}).}
  
  \bibitem[{\citenamefont{Caillol and
    Zeitlin}(2000)}]{Caillol_KineticEquationsStationary_2000}
  \bibinfo{author}{\bibfnamefont{P.}~\bibnamefont{Caillol}} \bibnamefont{and}
    \bibinfo{author}{\bibfnamefont{V.}~\bibnamefont{Zeitlin}},
    \bibinfo{journal}{Dynamics of Atmospheres and Oceans}
    \textbf{\bibinfo{volume}{32}}, \bibinfo{pages}{81} (\bibinfo{year}{2000}),
    ISSN \bibinfo{issn}{03770265}.
  
  \bibitem[{\citenamefont{Galtier}(2003)}]{Galtier_WeakInertialwaveTurbulence_2003}
  \bibinfo{author}{\bibfnamefont{S.}~\bibnamefont{Galtier}},
    \bibinfo{journal}{Physical Review E} \textbf{\bibinfo{volume}{68}},
    \bibinfo{pages}{015301} (\bibinfo{year}{2003}), ISSN
    \bibinfo{issn}{1063-651X, 1095-3787}.
  
  \bibitem[{\citenamefont{L'vov and
    Nazarenko}(2010)}]{Lvov_WeakTurbulenceKelvin_2010}
  \bibinfo{author}{\bibfnamefont{V.~S.} \bibnamefont{L'vov}} \bibnamefont{and}
    \bibinfo{author}{\bibfnamefont{S.}~\bibnamefont{Nazarenko}},
    \bibinfo{journal}{Low Temperature Physics} \textbf{\bibinfo{volume}{36}},
    \bibinfo{pages}{785} (\bibinfo{year}{2010}), ISSN \bibinfo{issn}{1063-777X,
    1090-6517}.
  
  \bibitem[{\citenamefont{D{\"u}ring et~al.}(2006)\citenamefont{D{\"u}ring,
    Josserand, and Rica}}]{during2006weak}
  \bibinfo{author}{\bibfnamefont{G.}~\bibnamefont{D{\"u}ring}},
    \bibinfo{author}{\bibfnamefont{C.}~\bibnamefont{Josserand}},
    \bibnamefont{and} \bibinfo{author}{\bibfnamefont{S.}~\bibnamefont{Rica}},
    \bibinfo{journal}{Physical review letters} \textbf{\bibinfo{volume}{97}},
    \bibinfo{pages}{025503} (\bibinfo{year}{2006}).
  
  \bibitem[{\citenamefont{Galtier and Nazarenko}(2017)}]{galtier2017turbulence}
  \bibinfo{author}{\bibfnamefont{S.}~\bibnamefont{Galtier}} \bibnamefont{and}
    \bibinfo{author}{\bibfnamefont{S.~V.} \bibnamefont{Nazarenko}},
    \bibinfo{journal}{Physical review letters} \textbf{\bibinfo{volume}{119}},
    \bibinfo{pages}{221101} (\bibinfo{year}{2017}).
  
  \bibitem[{\citenamefont{Dyachenko et~al.}(1992)\citenamefont{Dyachenko, Newell,
    Pushkarev, and Zakharov}}]{Dyachenko:1992aa}
  \bibinfo{author}{\bibfnamefont{S.}~\bibnamefont{Dyachenko}},
    \bibinfo{author}{\bibfnamefont{A.~C.} \bibnamefont{Newell}},
    \bibinfo{author}{\bibfnamefont{A.}~\bibnamefont{Pushkarev}},
    \bibnamefont{and} \bibinfo{author}{\bibfnamefont{V.~E.}
    \bibnamefont{Zakharov}}, \bibinfo{journal}{Physica D: Nonlinear Phenomena}
    \textbf{\bibinfo{volume}{57}}, \bibinfo{pages}{96} (\bibinfo{year}{1992}).
  
  \bibitem[{\citenamefont{Zakharov and Sagdeev}(1970)}]{ZakSag70}
  \bibinfo{author}{\bibfnamefont{V.~E.} \bibnamefont{Zakharov}} \bibnamefont{and}
    \bibinfo{author}{\bibfnamefont{R.~Z.} \bibnamefont{Sagdeev}},
    \bibinfo{journal}{Dokl. Akad. Nauk SSSR} \textbf{\bibinfo{volume}{192}},
    \bibinfo{pages}{297} (\bibinfo{year}{1970}).
  
  \bibitem[{\citenamefont{Kadomtsev and Petviashvili}(1973)}]{KadPet73}
  \bibinfo{author}{\bibfnamefont{B.~B.} \bibnamefont{Kadomtsev}}
    \bibnamefont{and} \bibinfo{author}{\bibfnamefont{V.~I.}
    \bibnamefont{Petviashvili}}, \bibinfo{journal}{Doklady Akademii Nauk SSSR}
    \textbf{\bibinfo{volume}{208}}, \bibinfo{pages}{794} (\bibinfo{year}{1973}).
  
  \bibitem[{\citenamefont{Newell and Aucoin}(1971)}]{newell_aucoin_1971}
  \bibinfo{author}{\bibfnamefont{A.~C.} \bibnamefont{Newell}} \bibnamefont{and}
    \bibinfo{author}{\bibfnamefont{P.~J.} \bibnamefont{Aucoin}},
    \bibinfo{journal}{Journal of Fluid Mechanics} \textbf{\bibinfo{volume}{49}},
    \bibinfo{pages}{593–609} (\bibinfo{year}{1971}).
  
  \bibitem[{\citenamefont{L'vov et~al.}(1997)\citenamefont{L'vov, L'vov, Newell,
    and Zakharov}}]{PhysRevE.56.390}
  \bibinfo{author}{\bibfnamefont{V.~S.} \bibnamefont{L'vov}},
    \bibinfo{author}{\bibfnamefont{Y.}~\bibnamefont{L'vov}},
    \bibinfo{author}{\bibfnamefont{A.~C.} \bibnamefont{Newell}},
    \bibnamefont{and} \bibinfo{author}{\bibfnamefont{V.}~\bibnamefont{Zakharov}},
    \bibinfo{journal}{Phys. Rev. E} \textbf{\bibinfo{volume}{56}},
    \bibinfo{pages}{390} (\bibinfo{year}{1997}).
  
  \bibitem[{\citenamefont{Navon et~al.}(2016)\citenamefont{Navon, Gaunt, Smith,
    and Hadzibabic}}]{Navon_EmergenceTurbulentCascade_2016}
  \bibinfo{author}{\bibfnamefont{N.}~\bibnamefont{Navon}},
    \bibinfo{author}{\bibfnamefont{A.~L.} \bibnamefont{Gaunt}},
    \bibinfo{author}{\bibfnamefont{R.~P.} \bibnamefont{Smith}}, \bibnamefont{and}
    \bibinfo{author}{\bibfnamefont{Z.}~\bibnamefont{Hadzibabic}},
    \bibinfo{journal}{Nature} \textbf{\bibinfo{volume}{539}}, \bibinfo{pages}{72}
    (\bibinfo{year}{2016}), ISSN \bibinfo{issn}{0028-0836, 1476-4687}.
  
  \bibitem[{\citenamefont{Navon et~al.}(2019)\citenamefont{Navon, Eigen, Zhang,
    Lopes, Gaunt, Fujimoto, Tsubota, Smith, and
    Hadzibabic}}]{Navon_SyntheticDissipationCascade_2019}
  \bibinfo{author}{\bibfnamefont{N.}~\bibnamefont{Navon}},
    \bibinfo{author}{\bibfnamefont{C.}~\bibnamefont{Eigen}},
    \bibinfo{author}{\bibfnamefont{J.}~\bibnamefont{Zhang}},
    \bibinfo{author}{\bibfnamefont{R.}~\bibnamefont{Lopes}},
    \bibinfo{author}{\bibfnamefont{A.~L.} \bibnamefont{Gaunt}},
    \bibinfo{author}{\bibfnamefont{K.}~\bibnamefont{Fujimoto}},
    \bibinfo{author}{\bibfnamefont{M.}~\bibnamefont{Tsubota}},
    \bibinfo{author}{\bibfnamefont{R.~P.} \bibnamefont{Smith}}, \bibnamefont{and}
    \bibinfo{author}{\bibfnamefont{Z.}~\bibnamefont{Hadzibabic}},
    \bibinfo{journal}{Science} \textbf{\bibinfo{volume}{366}},
    \bibinfo{pages}{382} (\bibinfo{year}{2019}), ISSN \bibinfo{issn}{0036-8075,
    1095-9203}.
  
  \bibitem[{\citenamefont{Johnstone et~al.}(2019)\citenamefont{Johnstone,
    Groszek, Starkey, Billington, Simula, and
    Helmerson}}]{Johnstone_EvolutionLargescaleFlow_2019}
  \bibinfo{author}{\bibfnamefont{S.~P.} \bibnamefont{Johnstone}},
    \bibinfo{author}{\bibfnamefont{A.~J.} \bibnamefont{Groszek}},
    \bibinfo{author}{\bibfnamefont{P.~T.} \bibnamefont{Starkey}},
    \bibinfo{author}{\bibfnamefont{C.~J.} \bibnamefont{Billington}},
    \bibinfo{author}{\bibfnamefont{T.~P.} \bibnamefont{Simula}},
    \bibnamefont{and}
    \bibinfo{author}{\bibfnamefont{K.}~\bibnamefont{Helmerson}},
    \bibinfo{journal}{Science} \textbf{\bibinfo{volume}{364}},
    \bibinfo{pages}{1267} (\bibinfo{year}{2019}), ISSN \bibinfo{issn}{0036-8075,
    1095-9203}.
  
  \bibitem[{\citenamefont{Gauthier et~al.}(2019)\citenamefont{Gauthier, Reeves,
    Yu, Bradley, Baker, Bell, {Rubinsztein-Dunlop}, Davis, and
    Neely}}]{Gauthier_GiantVortexClusters_2019}
  \bibinfo{author}{\bibfnamefont{G.}~\bibnamefont{Gauthier}},
    \bibinfo{author}{\bibfnamefont{M.~T.} \bibnamefont{Reeves}},
    \bibinfo{author}{\bibfnamefont{X.}~\bibnamefont{Yu}},
    \bibinfo{author}{\bibfnamefont{A.~S.} \bibnamefont{Bradley}},
    \bibinfo{author}{\bibfnamefont{M.~A.} \bibnamefont{Baker}},
    \bibinfo{author}{\bibfnamefont{T.~A.} \bibnamefont{Bell}},
    \bibinfo{author}{\bibfnamefont{H.}~\bibnamefont{{Rubinsztein-Dunlop}}},
    \bibinfo{author}{\bibfnamefont{M.~J.} \bibnamefont{Davis}}, \bibnamefont{and}
    \bibinfo{author}{\bibfnamefont{T.~W.} \bibnamefont{Neely}},
    \bibinfo{journal}{Science} \textbf{\bibinfo{volume}{364}},
    \bibinfo{pages}{1264} (\bibinfo{year}{2019}), ISSN \bibinfo{issn}{0036-8075,
    1095-9203}.
  
  \bibitem[{\citenamefont{Zakharov and Nazarenko}(2005)}]{Zakharov:2005aa}
  \bibinfo{author}{\bibfnamefont{V.~E.} \bibnamefont{Zakharov}} \bibnamefont{and}
    \bibinfo{author}{\bibfnamefont{S.~V.} \bibnamefont{Nazarenko}},
    \bibinfo{journal}{Physica D: Nonlinear Phenomena}
    \textbf{\bibinfo{volume}{201}}, \bibinfo{pages}{203} (\bibinfo{year}{2005}).
  
  \bibitem[{\citenamefont{M\"uller et~al.}(2021)\citenamefont{M\"uller, Polanco,
    and Krstulovic}}]{mullerIntermittencyVelocityCirculation2021}
  \bibinfo{author}{\bibfnamefont{N.~P.} \bibnamefont{M\"uller}},
    \bibinfo{author}{\bibfnamefont{J.~I.} \bibnamefont{Polanco}},
    \bibnamefont{and}
    \bibinfo{author}{\bibfnamefont{G.}~\bibnamefont{Krstulovic}},
    \bibinfo{journal}{Phys. Rev. X} \textbf{\bibinfo{volume}{11}},
    \bibinfo{pages}{011053} (\bibinfo{year}{2021}),
    \urlprefix\url{https://link.aps.org/doi/10.1103/PhysRevX.11.011053}.
  
  \bibitem[{\citenamefont{Krstulovic and
    Brachet}(2011)}]{krstulovicEnergyCascadeSmallscale2011}
  \bibinfo{author}{\bibfnamefont{G.}~\bibnamefont{Krstulovic}} \bibnamefont{and}
    \bibinfo{author}{\bibfnamefont{M.}~\bibnamefont{Brachet}},
    \bibinfo{journal}{Phys. Rev. E} \textbf{\bibinfo{volume}{83}},
    \bibinfo{pages}{066311} (\bibinfo{year}{2011}),
    \urlprefix\url{https://link.aps.org/doi/10.1103/PhysRevE.83.066311}.
  
  \bibitem[{\citenamefont{Nore et~al.}(1997)\citenamefont{Nore, Abid, and
    Brachet}}]{Nore97}
  \bibinfo{author}{\bibfnamefont{C.}~\bibnamefont{Nore}},
    \bibinfo{author}{\bibfnamefont{M.}~\bibnamefont{Abid}}, \bibnamefont{and}
    \bibinfo{author}{\bibfnamefont{M.~E.} \bibnamefont{Brachet}},
    \bibinfo{journal}{Phys. Rev. Lett.} \textbf{\bibinfo{volume}{78}},
    \bibinfo{pages}{3896} (\bibinfo{year}{1997}).
  
  \bibitem[{\citenamefont{Landau and Lifshitz}(1987)}]{Landau1987Fluid}
  \bibinfo{author}{\bibfnamefont{L.~D.} \bibnamefont{Landau}} \bibnamefont{and}
    \bibinfo{author}{\bibfnamefont{E.~M.} \bibnamefont{Lifshitz}},
    \emph{\bibinfo{title}{Fluid Mechanics, Second Edition: Volume 6 (Course of
    Theoretical Physics)}}, Course of theoretical physics / by L. D. Landau and
    E. M. Lifshitz, Vol. 6 (\bibinfo{publisher}{Butterworth-Heinemann},
    \bibinfo{year}{1987}), \bibinfo{edition}{2nd} ed., ISBN
    \bibinfo{isbn}{0750627670}.
  
  \end{thebibliography}
\end{document}